# Implementation of a level-set based volume penalization method for solving fluid flows around bluff bodies in OpenFOAM

Prashant Kumar[*], Vivek Kumar[*], Di Chen, Yosuke Hasegawa[#]
*Institute of Industrial Science, The University of Tokyo, Tokyo, 153-8505, Japan*
#Correspondence to: ysk@iis.u-tokyo.ac.jp
*These authors equally contributed.

A volume penalization-based immersed boundary technique is developed and thoroughly validated for fluid flow problems, specifically flow over bluff bodies. The proposed algorithm has been implemented in an Open Source Field Operation and Manipulation (OpenFOAM), a computational fluid dynamics solver. The immersed boundary method offers the advantage of inserting a complex solid object inside a Cartesian grid system, and therefore the governing equations can be applied to such a simpler grid arrangement. For capturing the fluid-solid interface more accurately, the grid is refined near the solid surface using topoSetDict and refineMeshDict utilities in OpenFOAM. In order to avoid any numerical oscillation and to compute the gradients accurately near the interface, the present volume penalization method (VPM) is integrated with a signed distance function, which is also referred to as a level-set function. Benchmark problems, such as flows around a cylinder and a sphere, are considered and thoroughly validated with the results available in the literature. For the flow over a stationary cylinder, the Reynolds number is varied so that it covers from a steady 2D (two-dimensional) flow to an unsteady 3D (three-dimensional) flow. The capability of the present solver has been further verified by considering the flow past a vibrating cylinder in the cross-stream direction. In addition, a flow over a sphere, which is inherently three-dimensional due to its geometrical shape, is validated in both steady and unsteady regimes. The results obtained by the present VPM show good agreement with those obtained by a body-fitted grid using the same numerical scheme as that of the VPM, and also with those reported in the literature. The present results indicate that the VPM-based immersed boundary technique can be widely applicable to scientific and engineering problems involving flow past stationary and moving bluff bodies of arbitrary geometry.

## I. INTRODUCTION

Flow past bluff bodies is a fundamental and crucial area of study in fluid mechanics and engineering. Bluff bodies induce complex flow phenomena such as flow separation and the formation of wake regions, and result in significant flow resistance. Understanding the fluid dynamics around bluff bodies is of significant importance in various engineering applications, such as aerospace, automotive, civil engineering, and environmental engineering. The study of flow past bluff bodies provides insights into aerodynamic drag, lift forces, heat transfer, and fluid-structure interactions. Numerical simulation is a powerful tool to reveal such complex flow phenomena and also to understand the underlying physics. Therefore, developing accurate and efficient numerical techniques dealing with complex geometries is crucial.

An immersed boundary method (IBM) has shown significant improvements in recent decades in examining the flow past stationary, moving as well as deformable bluff bodies. In this method, the conforming grid to the solid boundary is not required to capture the wake structure. Instead, the immersed boundary methods share a common characteristic of discretizing the governing equations for fluid flow on a conventional Cartesian grid. The equations for fluid flow and associated transport phenomena can be easily integrated and various solvers are readily available for such a simple grid. An arbitrary intricate geometry is then embedded into the computational domain. Since the numerical grid is generally not allocated on the interface, the boundary conditions at the fluid-solid interface are imposed by incorporating additional forcing terms into the equations. Various methods within this category encompass Lagrangian multipliers[3], level-set methods[4], fictitious domain approaches[1], and volume penalization approaches.[5,6]

In this study, we specifically focus on the volume penalization approach (VPM), which finds its physical basis in modeling solid surfaces as porous media with infinitesimal permeability. The mathematical justification for this approach can be found in Angot[7], and Carbou and Fabrie.[8] Notably, a no-slip boundary condition has been successfully implemented in various numerical experiments.[9-11] The method has demonstrated its capability in simulating various flow scenarios, including turbulent flows[11], incompressible viscous flows[12], compressible flows[13], and incompressible flows with scalar advection–diffusion.[14] Prodanovic and Bryant[15], for instance, utilized a Level-Set method to enforce a "no-penetration" constraint which effectively prevents a fluid from infiltrating the solid domain. Moreover, the VPM has been extended to accommodate not only the Dirichlet boundary condition[8] but also the Neumann and Robin conditions.[16]

The presence of a stationary object inside a fluid stream leads to complex flow phenomena. Moreover, it becomes even more complicated in the case of moving or deforming objects. Flows past cylinder[17-20] and sphere[21-24] are considered as typical classical problems where a blunt body is present in a fluid flow. The volume penalization method (VPM) offers a simplistic approach in which the object can be arbitrarily inserted in a Cartesian grid to examine the fluid-structure interaction (FSI). VPM has been previously implemented for FSI problems using spectral discretization.[25-27] Additionally, the Fourier pseudo-spectral method combined with VPM proved to be effective in investigating various aspects of insect flight, including the dynamics of leading-edge vortices[28] and the force distribution during takeoff.[29] The VPM is also employed by Kametani *et al.*[30] to develop a topology optimization framework for turbulent heat transfer surfaces.

OpenFOAM (Open Source Field Operation and Manipulation) has gained widespread usage in both academic research and industrial flow analysis.[31,32] It offers an efficient coding platform and a suitable environment for the users in the scientific community to implement and quickly disseminate new algorithms. In OpenFOAM, wall-boundary conditions are generally used for solid objects, while a recent implementation of the IBM based on the discrete forcing approach by Jasak *et al.*[33] has shown improvements in studying unsteady or deforming objects. A careful verification and validation have been carried out using this approach for different fluid flow problems considering the flow past cylinder and sphere.[34] More recently, VPM is applied in OpenFOAM for coupled radiative-conductive heat transfer problems by Ming and Hasegawa.[35]

To the best of our knowledge, however, the VPM method has not been implemented in OpenFOAM to disclose the physics involved in flows past bluff bodies. This leads to the prime motivation behind the present work to investigate how accurately the VPM can predict the flow behind the bluff bodies. For this purpose, we implement the VPM into OpenFOAM and thoroughly validate it in benchmark problems such as flow past a cylinder and a sphere. We also carry out additional simulations with body-fitted grids using the same numerical technique as that used in VPM for each case to verify the accuracy of the present VPM. The results obtained from the present VPM computations are also compared with those obtained by the previous discrete forcing approach[34] and also existing studies.

The present research study proceeds as follows: In Section II, the flow configurations considered in the present study, and also the governing equations and boundary conditions are introduced. Then, the treatment of solid objects inside a fluid domain using the VPM-based immersed boundary technique and its numerical implementation in OpenFOAM are explained. Section III presents the validation of the present scheme in flows past a cylinder and a sphere. Specifically, in Section IIIA, the results of 2D (two-dimensional) and 3D (three-dimensional) flow past cylinders with stationary and moving boundaries are presented. In Section IIIB, a flow past a sphere is considered for different values of the Reynolds number (Re). A particular focus is on the transition of the wake from steady axisymmetric to unsteady non-axisymmetric structures with increasing the Reynolds number. Finally, the conclusions drawn from the present study are summarized in Section IV.

## II. NUMERICAL METHODOLOGY

### A. Governing equations

In the present study, a fluid flow is assumed to be incompressible, laminar, and unsteady. The governing equations for the mass and momentum conservations in the non-dimensionalized form can be expressed as

$$\frac{\partial u_i}{\partial x_i} = 0, \qquad (1)$$

$$\frac{\partial u_i}{\partial t} + \frac{\partial (u_j u_i)}{\partial x_j} = -\frac{\partial p}{\partial x_i} + \frac{1}{Re}\frac{\partial^2 u_i}{\partial x_j \partial x_j} - \eta_u \phi (u_i - u_i^s). \qquad (2)$$

Here $u_i$ (= $u, v, w$ for $i$ = 1, 2, 3, respectively) are the velocity components along the $x_i$ (= $x, y, z$ for $i$ = 1, 2, 3, respectively) directions, whereas $p$ is the static pressure. The upstream velocity ($U_\infty$) and the diameter of a cylinder or sphere ($D$) are considered as the characteristic velocity- and characteristic length-scales, respectively, so that the velocity components and the space coordinates are non-dimensionalized using $U_\infty$ and $D$. The static pressure is normalized by $\rho U_\infty^2$. The Reynolds number (Re) is defined as Re $= \rho U_\infty D/\mu$, where $\rho$ and $\mu$ stand for the density and the dynamics viscosity of the fluid, respectively.

The present numerical approach is based on the volume penalization method (VPM), in which a solid surface is embedded in a Cartesian grid system. This method relies on the fact that solid objects can be treated as porous media with zero permeability. In order to represent the presence of the solid object in the computational domain, the last term in Eqn. (2) is added as an artificial body force to impose a no-slip boundary condition on a solid surface. The body force is null in the fluid region, whereas it is forced to be non-zero to meet the target velocity of the solid region ($u_i^s$). The solid and fluid regions are differentiated by a phase function $\phi$, which possesses 1 and 0 values in the solid and fluid regions, respectively. A more detailed definition of $\phi$ will be explained in the next subsection. $\eta_u$ is a penalization coefficient and is set to be sufficiently large ($\eta_u = 10^6$) to satisfy the no-slip boundary condition and to maintain numerical stability.

**B. Expressing solid-fluid interface (An immersed boundary approach)**

In the present immersed boundary approach, the complex solid-fluid interface is represented by a signed distance function, which is often referred to as a level-set function $\phi_0$.[36,37] An arbitrary three-dimensional solid-fluid interface corresponds to the zero iso-surface of $\phi_0$, while it is defined to be negative and positive in the fluid and solid regions, respectively. The distribution of $\phi_0$ for an arbitrary solid region is depicted in Fig. 1(a). Once the level-set function $\phi_0$ is defined, we can define a phase indicator $\phi$, which is zero and unity in the fluid and solid regions, respectively, as follows:

$$\phi = \begin{cases} 0 & \phi_0 < -\delta_t \\ \frac{1}{2}\left[\sin\left(\frac{\pi \phi_0}{2\delta_t}\right) + 1\right] & -\delta_t \leq \phi_0 \leq \delta_t \\ 1 & \phi_0 > \delta_t. \end{cases} \quad (3)$$

Note that $\phi$ changes smoothly from 0 to 1 in the form of a sine function across the solid-fluid interface. Ideally, the fluid-solid interface should be infinitesimally thin, and the continuous and smooth representation (3) of $\phi$ is introduced to reduce numerical instability. Here, $\delta_t$ is the one-sided transition layer thickness and is usually defined as $\delta_t = K_\delta \Delta x_{int}$, where $\Delta x_{int}$ is the local grid spacing at the fluid-solid interface and $K_\delta$ is a multiplication factor. In the present study, $K_\delta$ is fixed as 1.5. Generally, this region is considered as 1 to 2 grid size on each side of the interface. The spatial distribution of $\phi$ near the fluid-solid interface is illustrated in Fig. 1(b).

An initial shape is embedded in the rectangular grid and the solid and fluid regions are converted in the form of $\phi_0$. For $\phi_0$ to be a signed distance function, it has to satisfy the Eikonal equation $|\nabla \phi_0| = 1$ throughout the computational domain.[36,37] In order to achieve this, we solve the following equation for the pseudo-time, known as the reinitialization of the level-set function ($\phi_0$).

$$\frac{\partial \phi_0}{\partial \tau} + (\widetilde{\boldsymbol{u}} \cdot \nabla)\phi_0 = \text{sign}(\phi_0) + D\nabla^2 \phi_0, \quad (4)$$

Here $\tau$ is the pseudo time, $\widetilde{\boldsymbol{u}} = \text{sign}(\phi_0)\nabla \phi_0/|\nabla \phi_0|$ is the wall-normal velocity and $D$ is an artificial diffusion coefficient to reduce numerical error. The value of $D$ is generally taken as half of the grid size near the interface to make sure that the diffusion term is not dominant. The distribution of $\phi_0$ is shown in Fig. 2(a) for a solid object of circular shape inserted in the fluid region. The yellow and red curves indicate the positive and negative values of $\phi_0$ near the interface. The interface is located as $\phi_0 = 0$ in the green color. The contours in Fig. 2(a) reveal the spatial distribution of $\phi$. It smoothly varies from 0 in the fluid region to 1 in the solid region. In order to confirm that the interface is accurately captured after the reinitialization, the original as well as the reconstructed interface (given by $\phi_0 = 0$) is illustrated in Fig. 2(b). The interface indicated by $\phi_0 = 0$ overlaps the original interface. It also confirms that the present grid is sufficiently refined near the solid-fluid interface. Note that, for simple geometries such as a cylinder or a sphere considered in the present study, it is not necessary to conduct the reinitialization procedure (4), since the analytical spatial distributions of the level-set functions for them are readily available. However, the reinitialization is generally required to represent non-trivial geometries by the level-set method. The present results demonstrate the robustness of the present reinitialization procedure for constructing the level-set function.

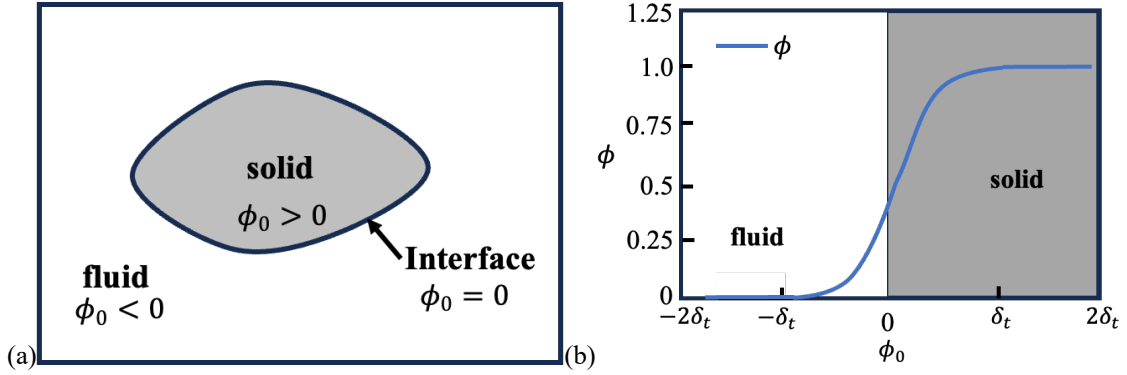

**FIG. 1.** Spatial distribution of (a) level-set function ($\phi_0$) and (b) phase indicator ($\phi$) as a function of $\phi_0$.

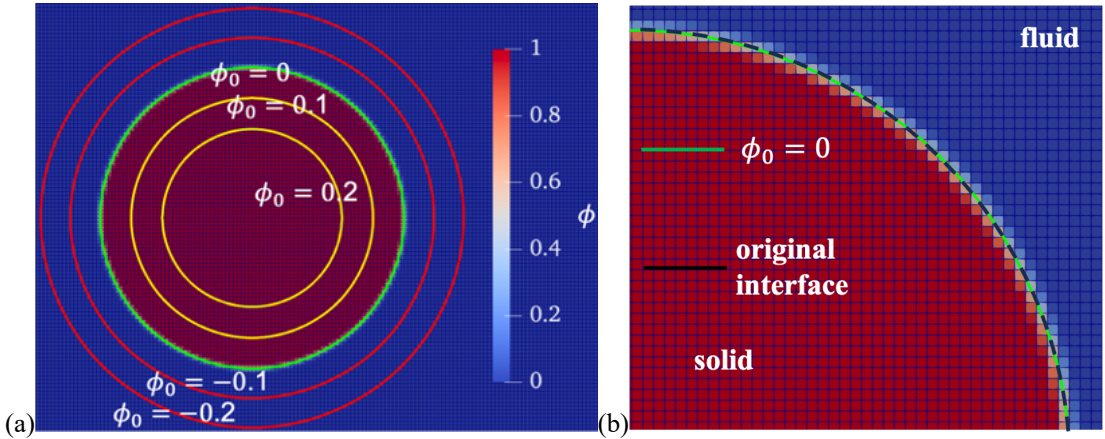

**FIG. 2.** (a) Distribution of $\phi$ and $\phi_0$ for a solid object of circular shape inserted in the fluid domain. [$\phi_0 = 0$ (green curve): reconstructed interface, yellow and red curves indicate positive and negative values of $\phi_0$] (b) the original and reconstructed interfaces [black curve: original interface, green curve: reconstructed interface].

## C. Implementation in OpenFOAM

A new solver based on VPM has been developed in OpenFOAM C++ library. The three-dimensional unsteady continuity and momentum equations are discretized on finite-volume hexahedral Cartesian grids. The pressure Implicit with the Splitting of Operators (PISO) algorithm is employed to resolve the coupling between pressure and velocity. The divergence, gradient, and Laplacian terms of the governing equations are discretized using the standard Gaussian finite-volume integration method. These terms have been discretized using a second-order linear scheme. A second-order backward implicit scheme is employed for the time derivative in the momentum equation. Absolute convergence criteria for both pressure and velocity are fixed as $10^{-6}$. In order to maintain accuracy and numerical stability, the maximum Courant number is set as 0.6.

Fig.3 shows the flow chart of the present numerical implementation used for the stationary or moving objects in OpenFOAM. The complete procedure can be divided into four steps as outlined below.

**Step1:** We start the numerical procedure by generating a Cartesian grid in a prescribed rectangular computational domain using blockMesh in OpenFoam. The grid is refined up to $3^{rd}$ or $4^{th}$ level based on the requirement around the solid object using topoSetDict and refineMeshdict utilities in OpenFOAM. Firstly, the refinement region is selected using topoSetDict utility and then the grid is refined in that region using refineMeshDict. In each refinement, the grid is halved of the previous grid size.

**Step2:** In this procedure, the solid and fluid regions are separated. We insert a solid object in the form of an STL file in the refined Cartesian mesh generated in the previous step. A phase indicator function ($\phi$) is initialized based on the STL file and each point in the domain is converted into the binary field 0 (for fluid) or 1 (for solid). An initial $\phi_0$ is generated based on the binary field. The $\phi_0$ is reinitialized using Eqn. (4) to maintain the signed distance property. Then $\phi$ is expressed in terms of using Eqn. (3). Then, the different terms in the governing equations are discretized as discussed above.

**Step3:** For a flow over a stationary object, the governing equations are solved using the above discretization schemes. The discretized equations are solved till the final converged results are obtained.

**Step4:** In the case of a moving object, the governing equations are first solved for flow past a stationary object till the flow reaches a dynamically steady state. The final converged flow field over a stationary object is used as an initial condition for flow past a moving object. This is done in order to reduce any diffusion of the phase indicator during the movement of the solid object in the initial transient development of the flow field. Then the motion of the object is started. In each iteration, the solid object is shifted to a new location, and then governing equations are solved till the final convergence. The shifting of solid object is done by updating the phase indicator to a new location as time progresses.

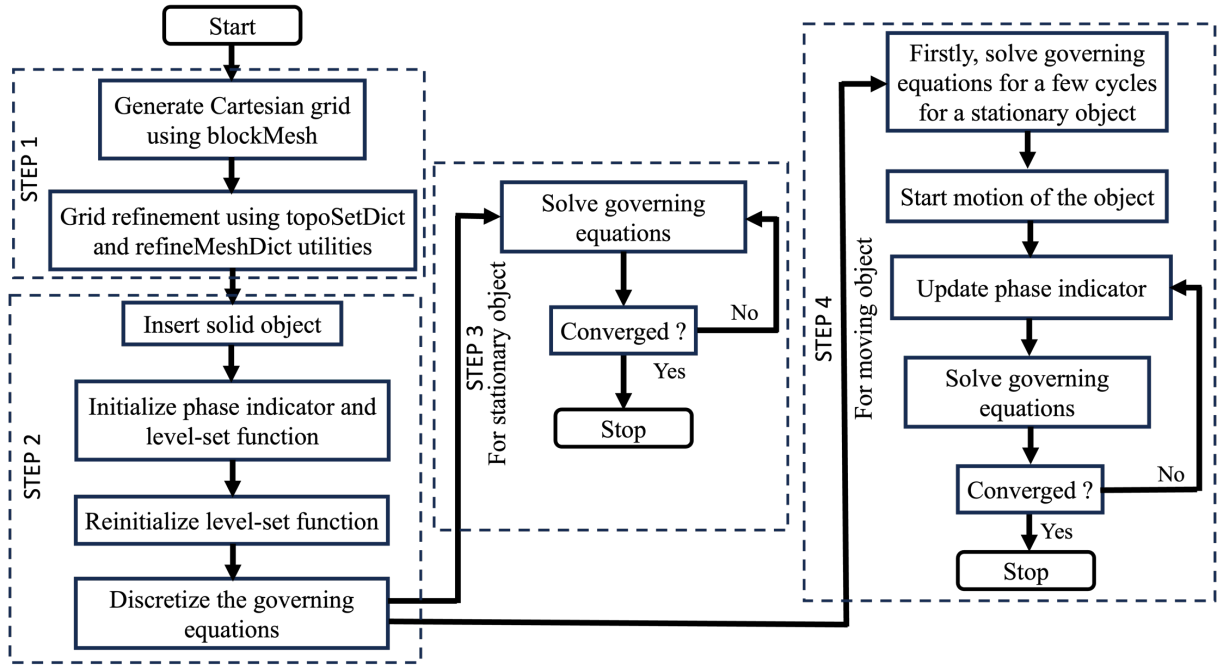

**FIG. 3.** Flow chart of the numerical algorithm used for stationary or moving objects.

### III. VALIDATION OF THE SOLVER

#### A. 2D and 3D flows past circular cylinders

*1. Computational domain and grid for* **2D** *flow past a* **circular cylinder**

At low Re (< 190), the flow is reported to be 2D flow.[17,18,38,42] In such a low Re regime, the flow can be steady or unsteady 2D. In the present study, the flow is simulated for both 2D steady and 2D unsteady flow in order to demonstrate the robustness of the algorithm. Figure 4 shows the schematic of the computational domain for flow past a circular cylinder. For a flow past an infinitely long cylinder, i.e., 2D case, the dimensions of the computational domain in $x$, and $y$-directions are $L_x$ (= $L_U$ + $L_D$) = 48$D$ and $L_y$ = 32$D$, respectively. Here, $D$ is the diameter of the cylinder and is assumed to be the characteristic dimension for flow past a circular cylinder. The cylinder is located at a distance of 16$D$ from the inlet boundary. The dimensions are considered to be similar as reported in the previous studies.[17,34] The origin of the coordinate system is fixed at the centerline of the circular cylinder.

The mass and momentum conservation equations have been solved by employing proper boundary conditions on different surfaces of the computational domain.

*Inlet*: Uniform velocity ($u = 1$, $v = 0$, $w = 0$).
*Outlet*: Pressure outlet ($p = 0$).
*Top and bottom surfaces*: Free-slip and impermeable boundary conditions.
$$v = 0, \frac{\partial u}{\partial y} = 0, \frac{\partial w}{\partial y} = 0, \frac{\partial p}{\partial y} = 0 \ .$$
For 3D flow over a cylinder, the periodic boundary condition is employed on the side surfaces, i.e., in the spanwise direction (z-direction).

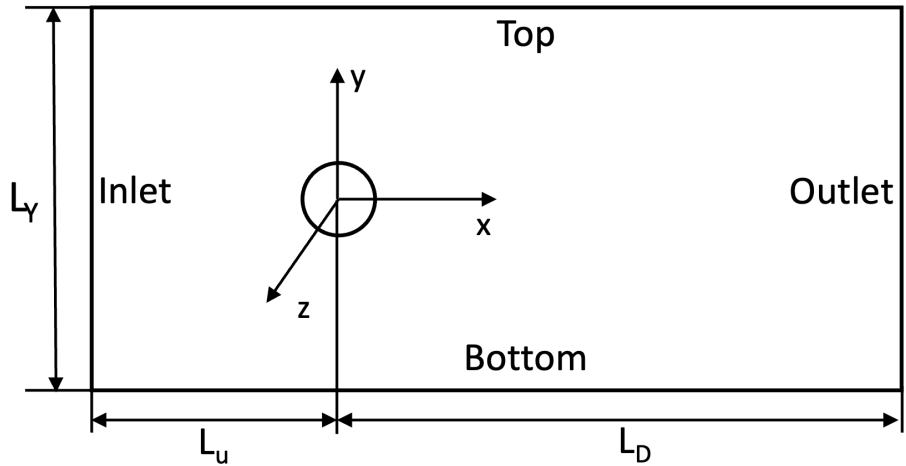

**FIG. 4.** Schematic of the two-dimensional computational domain.

Figure 5 depicts the grid distribution in the computational domain. The rectangular grid with different levels of grid refinement is generated for the VPM case. The refinement is carried out using topoSetDict utility in the OpenFOAM, in which the region is selected for the refinement. The grid is refined up to $4^{th}$ level, as shown in Fig. 5(a). For the body-fitted case, the structured hexahedral grid is generated in the whole computational domain with high grid density around the cylinder using ANSYS ICEM CFD. The grid is a non-uniform O-grid and is stretched with a constant stretching ratio of less than 1.1, as illustrated in Fig. 5(b).

A thorough grid independence study has been conducted for both VPM and body-fitted cases at Re = 100. For the VPM case, five different levels of the grid are tested and the corresponding mean drag coefficients ($Cd_{mean}$) and root-mean-square (RMS) values of the lift coefficient ($Cl_{rms}$) are reported in Table I(a). Firstly $0^{th}$ level uniform mesh is generated considering 240×160 grid divisions in *x* and *y*-directions, respectively. The region of refinement (non-dimensionalized with the characteristic dimension) for each level is also included in Table I(a). In each refinement, the mesh size becomes half of the previous size. It can be clearly confirmed that both $Cd_{mean}$ and $Cl_{rms}$ converge with the grid refinement. Specifically, the grid dependency becomes negligible as the grid is refined from level 4 to level 5. Therefore, all the computations are performed considering $4^{th}$ level of grid refinement for the VPM case.

For the body-fitted case, the grid distribution is varied in both the radial and circumferential directions for the grid convergence test. The values of $Cd_{mean}$ and $Cl_{rms}$ are also presented in Table I(b) for four different grid arrangements around the cylinder. The grid dependency of $Cd_{mean}$ and $Cl_{rms}$ monotonically decreases as increasing the grid resolution near the cylinder, and the values of $Cd_{mean}$ and $Cl_{rms}$ are almost converged as the number of cells increases from 200 to 240 and the grid size near the wall reduces from 0.005 to 0.002, as shown in Table I(b). Therefore, further computations are carried out with the number of cells around the cylinder of 200 which corresponds to the near-wall grid size of 0.005.

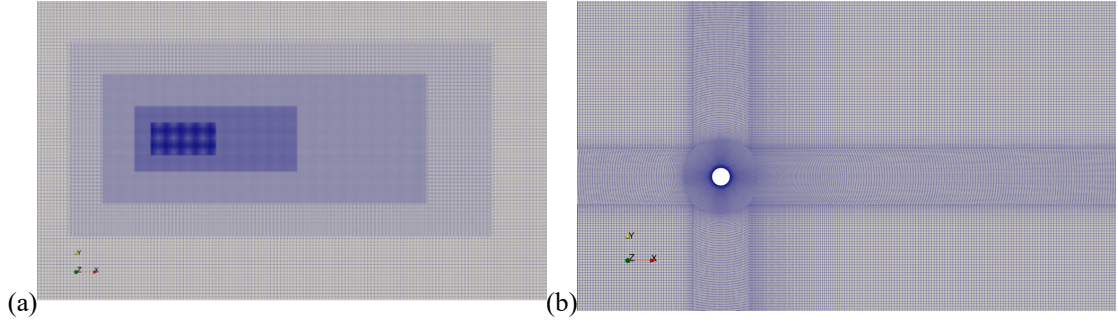

(a)  (b)

**FIG. 5.** Grid distribution around cylinder (a) VPM and (b) body-fitted.

**TABLE I(a).** Grid independence study for VPM case at Re = 100.

| Refinement | Region ($x_{min}$ $y_{min}$ $z_{min}$) to ($x_{max}$ $y_{max}$ $z_{max}$) | $Cd_{mean}$ | % Change in $Cd_{mean}$ | $Cl_{rms}$ | % change in $Cl_{rms}$ |
|---|---|---|---|---|---|
| 1st level | (-6 -6 0) to (20 6 1) | 1.33 | | 0.438 | |
| 2nd level | (-4 -4 0) to (16 4 1) | 1.35 | 1.50 | 0.456 | 4.11 |
| 3rd level | (-2 -2 0) to (8 2 1) | 1.37 | 1.48 | 0.464 | 1.75 |
| **4th level** | **(-1 -1 0) to (3 1 1)** | **1.38** | **0.73** | **0.468** | **0.86** |
| 5th level | (-0.6 -0.6 0) to (2 0.6 1) | 1.38 | 0 | 0.470 | 0.43 |

**TABLE I(b).** Grid independence study for body-fitted case at Re = 100.

| Number of cells around cylinder | Near-wall cell size | $Cd_{mean}$ | % Change in $Cd_{mean}$ | $Cl_{rms}$ | % change in $Cl_{rms}$ |
|---|---|---|---|---|---|
| 120 | 0.012 | 1.362 | | 0.484 | |
| 160 | 0.009 | 1.376 | 1.03 | 0.476 | 1.65 |
| **200** | **0.005** | **1.382** | **0.44** | **0.474** | **0.42** |
| 240 | 0.002 | 1.383 | 0.07 | 0.474 | 0 |

*2. Steady 2D flow past a circular cylinder*

At Re = 30, a steady flow with the formation of two symmetric wake bubbles behind the cylinder is obtained in both the VPM and body-fitted cases. Table II shows the comparison of different parameters in the wake and the mean drag coefficient on the cylinder ($Cd_{mean}$). The values obtained from the present VPM and body-fitted cases are in good agreement with the values reported in the literature both experimentally and numerically.

*3. Unsteady 2D flow past a circular cylinder*

The transition from a steady to unsteady flow occurs around Re = 40 for flow past a circular cylinder.[20,41] In order to assess the robustness of the present VPM solver, the unsteady nature of the wake has been simulated at Re = 100 and 185. Both mean ($Cd_{mean}$, $Cl_{rms}$, $\theta_s$) and temporal (Strouhal frequency: $St = fD/U_\infty$, where $f$ is the vortex shedding frequency) quantities are compared, as depicted in Table III. The values, summarized below, are in the same range as reported in the literature. The well-known wake structure, i.e., the Karman vortex street, at Re = 185 is depicted in Fig. 6 for both VPM and body-fitted cases. The vorticity contours indicate the periodic shedding of vortices which diffuse as they convect downstream. The spatial arrangement of vortices obtained from the VPM is quite similar to that of a body-fitted case. The topological shape of the vortices is also found to be similar to those reported in the previous studies[17,34,40] at the same Reynolds number of 185. The temporal evolutions of $Cl$ and $Cd$ are shown in Figs. 7(a) and 7(b) at the Reynolds numbers of 100 and 185. It can be clearly seen that, with increasing Re, the amplitudes of their fluctuations increase indicating the enhancement of the wake unsteadiness. Such trends have also been reported by Gilmanian and Queutey[40] and Constant *et al*.[34] The signals obtained by the preset VPM match quite closely with those obtained by the body-fitted mesh. Figure 8 illustrates the coefficient of pressure ($C_p = \frac{p-p_\infty}{\frac{1}{2}\rho U_\infty^2}$) distribution around the circumference of the cylinder at Re = 100. The pressure distribution from both VPM and body-fitted cases is in good agreement with the variation of $C_p$ observed by Qu et al.[41]

**TABLE II.** Comparison of results at Re = 30. **L**: length of the recirculation region; **a**: the distance between the cylinder rear surface and center of the recirculation region; **b**: the transverse distance between the centers of the two recirculation bubbles; $\boldsymbol{\theta_s}$: the separation angle from the rear stagnation point; **Cd**$_{mean}$: mean drag coefficient.

| Authors | Technique | L/D | a/D | b/D | $\theta_s$ | Cd$_{mean}$ |
|---|---|---|---|---|---|---|
| Pinelli et al. (2010)[17] | Simulation | 1.70 | 0.56 | 0.52 | 48.05 | 1.80 |
| Blackburn and Henderson (1999)[38] | Simulation | -- | -- | -- | -- | 1.74 |
| Coutanceau and Bouard (1977)[18] | Experiment | 1.55 | 0.54 | 0.54 | 50.00 | -- |
| Tritton (1959)[19] | Experiment | -- | -- | -- | -- | 1.74 |
| Constant et al. (2017)[34] | Simulation (IBM OpenFOAM) | 1.64 | 0.55 | 0.53 | 48.40 | 1.77 |
| Present | Simulation (body-fitted) | 1.65 | 0.54 | 0.52 | 48.01 | 1.81 |
| Present | Simulation (VPM) | 1.63 | 0.53 | 0.52 | 48.65 | 1.80 |

**TABLE III.** Comparison of results at Re = 100 and 185.

| | Authors | Technique | Cd$_{mean}$ | Cl$_{rms}$ | St | $\theta_s$ |
|---|---|---|---|---|---|---|
| Re = 100 | Blackburn and Henderson (1999)[38] | Simulation | 1.35 | -- | -- | -- |
| | Williamson (1996)[42] | Experiment | -- | -- | 0.164 | -- |
| | Henderson (1995)[43] | Simulation | 1.35 | -- | -- | -- |
| | Constant et al. (2017)[34] | Simulation (IBM OpenFOAM) | 1.37 | -- | 0.165 | 118.9 |
| | Present | Simulation (body-fitted) | 1.38 | 0.476 | 0.165 | 117.5 |
| | Present | Simulation (VPM) | 1.37 | 0.472 | 0.165 | 119.8 |
| Re = 185 | Vanella and Balaras (2009)[44] | Simulation | 1.377 | 0.461 | -- | -- |
| | Gilmanian and Queutey (2002)[40] | Simulation | 1.287 | 0.443 | 0.195 | -- |
| | Williamson (1988)[20] | Experiment | -- | -- | 0.193 | -- |
| | Constant et al. (2017)[34] | Simulation (IBM OpenFOAM) | 1.379 | 0.427 | 0.198 | 110.8 |
| | Present | Simulation (body-fitted) | 1.369 | 0.460 | 0.197 | 112.6 |
| | Present | Simulation (VPM) | 1.377 | 0.468 | 0.196 | 111.9 |

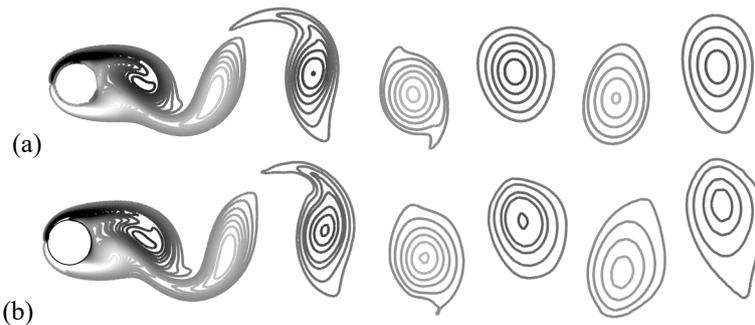

**FIG. 6.** Spanwise-vorticity (z-vorticity) contours indicating the Karman vortex street at Re = 185 (a) VPM and (b) body-fitted. Here z-vorticity magnitude varies from −1 (black) to 1 (gray).

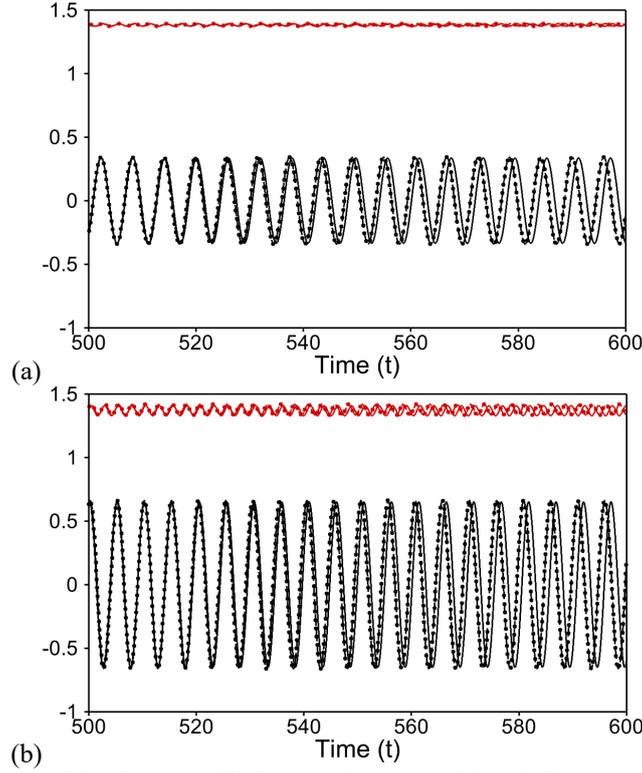

**FIG. 7.** Temporal evolutions of the force coefficients at (a) Re = 100 and (b) Re = 185 [solid line: body-fitted, dash line: VPM; black color: *Cl*, red color: *Cd*].

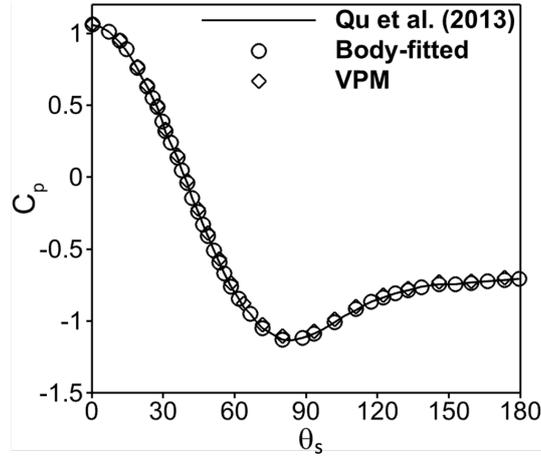

**FIG. 8.** Comparison of pressure distribution around the circumference of the cylinder at Re = 100.

### 4. Unsteady 2D flow past an oscillating circular cylinder

The capability of the present level-set-based VPM solver has been further assessed by considering the flow past a 2D oscillating circular cylinder at Re = 500. The computational domain and boundary conditions are the same as considered for the flow past a 2D stationary cylinder in the previous case. Following the previous study by Blackburn and Henderson[38], the cylinder is forced to oscillate in the cross-stream direction with a fixed value of amplitude and oscillation frequency. The amplitude ($a^* = a/D$) and frequency ($f^* = f/f_o$, $f_o$ is the vortex shedding frequency for the stationary cylinder at Re = 500) of the cylinder oscillation are 0.25 and 0.975, respectively. In order to avoid possible numerical instability, we first simulated the flow past a stationary cylinder at Re = 500. Once the flow reaches a dynamically steady state after several periods of vortex shedding, the cylinder is forced to oscillate in the cross-stream direction.

Figure 9 illustrates the vorticity contours at five different instants over half of the shedding cycle starting from the topmost position of the cylinder. Both VPM and body-fitted results are compared with those reported in the literature. It can be observed that the topological features of the vortices from the VPM at different instants closely resemble with those of the body-fitted case. In addition, the shear layers and vortex structures obtained from the present VPM and body-fitted cases match quite closely with the results reported in the literature.[34,38] The spatial arrangement of vortices as well as their temporal evolutions are in good agreement with the literature[34,38] corresponding to different instants of the cylinder oscillation. Figure 10 shows the cylinder response (lift coefficient) as a function of the cross-stream displacement of the cylinder. Again, the limit-cycle oscillation matches quite closely with Blackburn and Henderson.[38]

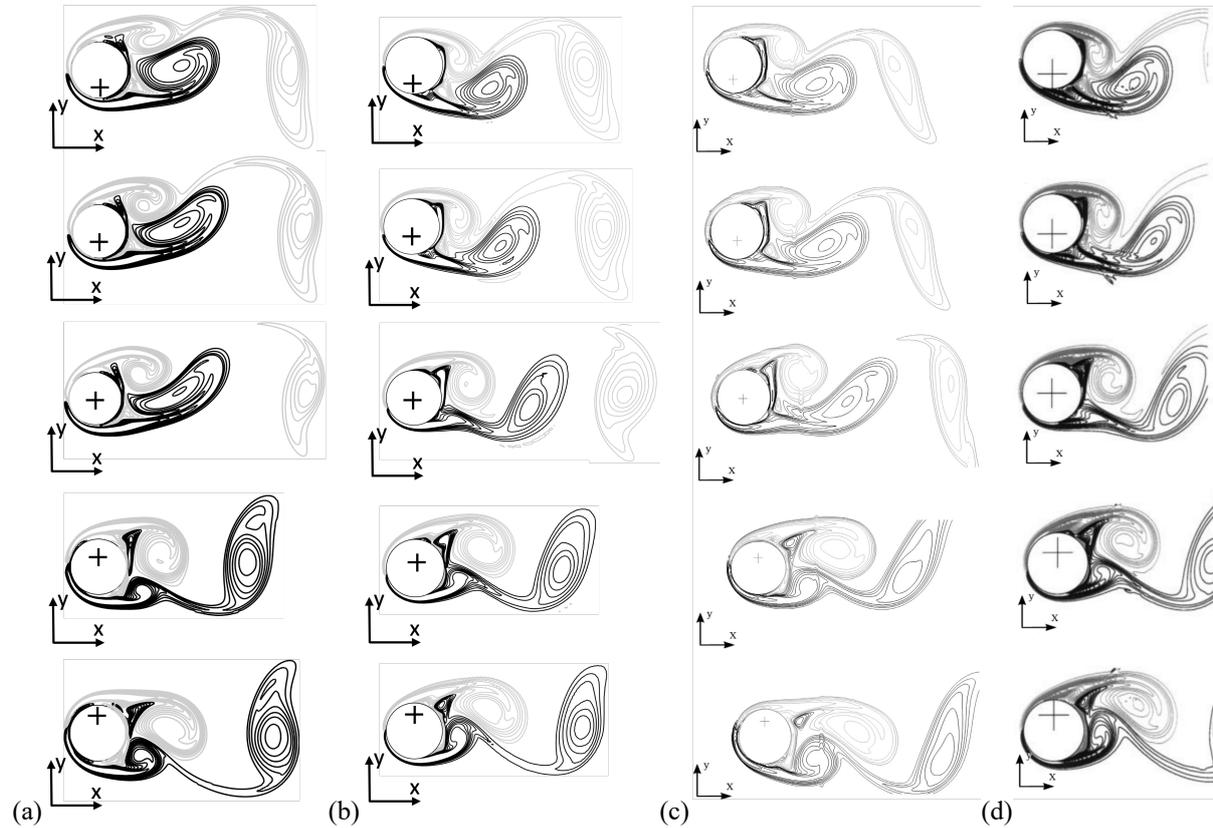

**FIG. 9.** Instantaneous vorticity contours for the flow past a 2D oscillating cylinder at Re = 500. (a) VPM, (b) body-fitted, (c) Num.[34] (IBM OpenFOAM), and (d) Blackburn and Henderson.[38] The contours are presented corresponding to the five different instants over half of the shedding cycle. Here $z$-vorticity magnitude varies from −1 (gray) to 1 (black).

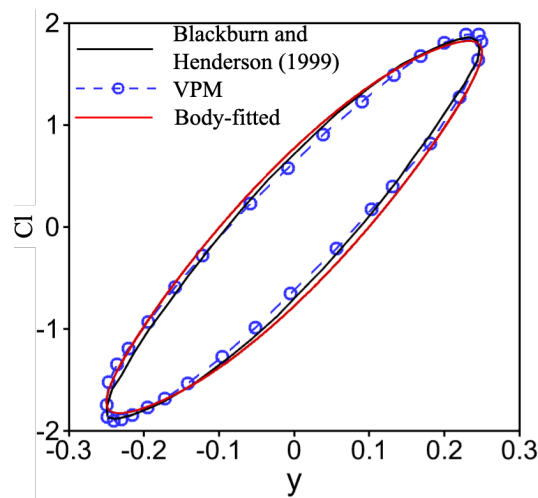

**FIG. 10.** Cylinder response ($Cl$) as a function of cross-stream displacement ($y$) for the flow past a 2D oscillating cylinder at Re = 500.

*5. Unsteady 3D flow past a circular cylinder*

For solving unsteady 3D flow past a cylinder, the dimensions of the computational domain in *x*, *y* and *z*-directions are set as $L_x (= L_U + L_D) = 48D$, $L_y = 32D$ and $L_z = 10.24D$, respectively. These dimensions are decided based on the previous study.[34] The cylinder is located at a distance of $L_U = 16D$ from the inlet boundary. Similar grid arrangement has been considered to that used for the 2D flows discussed in the previous subsections. In the present case, the grid is also refined along the axis of the cylinder in order to accurately capture the 3D transitions.

The transition from 2D to 3D unsteady flow occurs around Re = 190.[42] This transition corresponds to mode A and mode B instabilities. In order to capture the 3D transitional flow regime, computations are carried out at Re = 200 and 300 using both VPM and body-fitted grids. In Table IV, the values of $Cd_{mean}$, $Cl_{rms}$, and St obtained in the present simulations are compared with experimental and numerical results reported in the literature. Both the time averaged and fluctuating quantities observed from the present simulations agree well with those reported in the literature. In addition, the present VPM results are in good agreement with the present body-fitted results at each Re. The temporal signals are found to be of the harmonic type for the flow past a 2D cylinder, as depicted in Fig. 7. The transition from unsteady 2D to unsteady 3D perturbs the periodic arrangement of vortices. Modulations in the signals can be clearly seen at Re = 200 and 300, as illustrated in Fig. 11. It can also be observed that the amplitude of the fluctuation increases with increasing Re. These results were obtained from the present VPM, and similar variations were also observed for the present body-fitted simulation (not shown here). Figure 12 shows the pressure distribution around the circumference of the cylinder at Re = 200. The pressure variation also shows excellent agreement with the data reported by Qu *et al.*[41] and Norberg.[47]

**TABLE IV.** Comparison of results at Re = 200 and 300.

|  | **Authors** | **Technique** | *Cd<sub>mean</sub>* | *Cl<sub>rms</sub>* | **St** |
|---|---|---|---|---|---|
| Re = 200 | Rajani *et al.* (2009)[45] | Simulation | 1.338 | 0.4216 | 0.1936 |
|  | Qu *et al.* (2013)[41] | Simulation | 1.240 | 0.339 | 0.1801 |
|  | Williamson (1996)[42] | Experiment | -- | -- | 0.1800 |
|  | Constant *et al.* (2017)[34] | Simulation (IBM OpenFOAM) | 1.384 | 0.346 | 0.1802 |
|  | Present | Simulation (body-fitted) | 1.317 | 0.408 | 0.1803 |
|  | Present | Simulation (VPM) | 1.351 | 0.360 | 0.1805 |
| Re = 300 | Rajani *et al.* (2009)[45] | Simulation | 1.28 | 0.499 | 0.195 |
|  | Mittal and Balachandar (1995)[46] | Simulation | 1.26 | 0.380 | 0.203 |
|  | Williamson (1996)[42] | Experiment | -- | -- | 0.203 |
|  | Norberg (1993)[47] | Experiment | -- | 0.435 | 0.203 |
|  | Wieselsberger (1922)[48] | Experiment | 1.22 | -- | -- |
|  | Constant *et al.* (2017)[34] | Simulation (IBM OpenFOAM) | 1.43 | 0.453 | 0.198 |
|  | Present | Simulation (body-fitted) | 1.34 | 0.457 | 0.196 |
|  | Present | Simulation (VPM) | 1.34 | 0.469 | 0.196 |

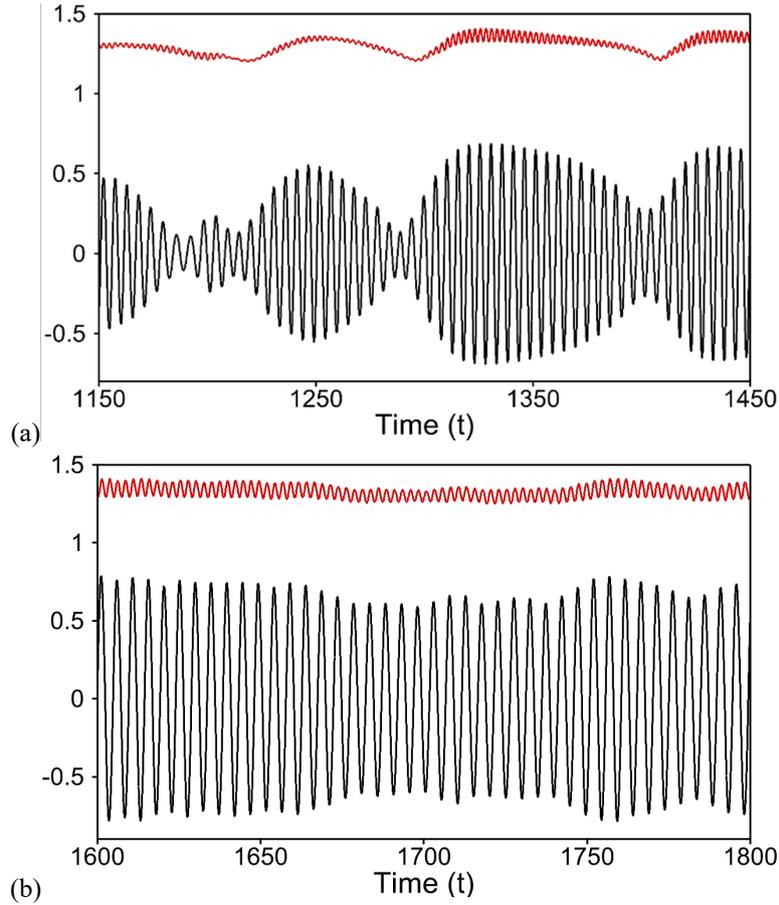

**FIG. 11.** The temporal evolutions of the force coefficients at (a) Re = 200 and (b) Re = 300 using VPM [red line: *Cd*, black line: *Cl*].

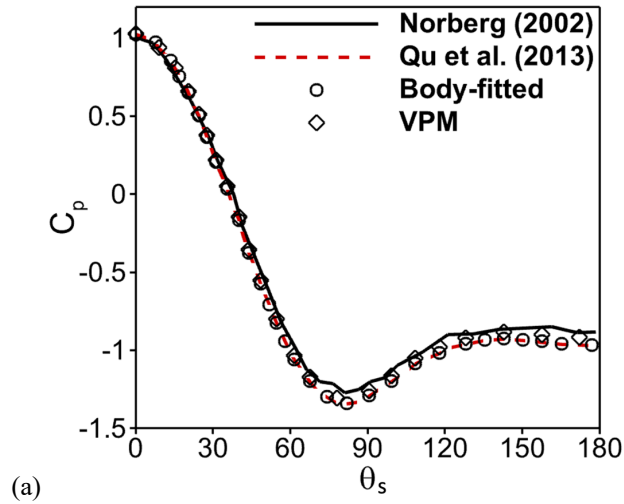

**FIG. 12.** Comparison of pressure distribution around the circumference of the cylinder at Re = 200.

### B. Flow past a sphere
#### *1. Computational domain and grid*

In order to further assess the accuracy of the present VPM, the three-dimensional flow around a sphere that consists of a complex wake structure has been considered. The Reynolds number (Re) based on the diameter of the sphere ($D$) and free-stream velocity ($U_\infty$) is varied from 100 to 300. In this regime of Re, the wake flow behavior changes as steady axisymmetry flow, steady non-axisymmetry flow, and finally to unsteady non-axisymmetry flow.[21] Different local as well as global parameters are compared with the results from the literature.

The sphere is placed at a distance of 16$D$ from the inlet. The downstream boundary is 32$D$ from the center of the sphere. The origin of the coordinate system is located at the center of the sphere. The side surfaces are 16$D$ apart from the origin of the coordinate system. The dimensions of the computational domain are similar to the dimensions considered by Constant *et al*.[34] The details of the domain dimensions and the grid distribution around the sphere are illustrated in Fig. 13. For the body-fitted case, similar to the flow past a circular cylinder, the grid is refined around the sphere and gradually stretched away from the surface of the sphere with a stretching ratio of less than 1.1, as depicted in Fig. 13(a). For the VPM case, firstly uniform grid (0-level grid), i.e., $\Delta x = \Delta y = \Delta z = 0.2$ (non-dimensionalized by the diameter of the sphere) is generated in the whole computational domain. The fourth level of grid refinement is carried out using the topoSetDict and refineMeshDict utilities in OpenFOAM to accurately capture the wake dynamics around the flow past a sphere. In each refinement, the grid is halved of the previous resolution in all three dimensions, as shown in Fig. 13(b).

The boundary conditions on different surfaces of the computational domain are as follows:

*Inlet*: Uniform velocity ($u = 1$, $v = 0$, $w = 0$).

*Outlet*: Pressure outlet ($p = 0$).

*Top, bottom, and side surfaces*: Free-slip and impermeable boundary conditions.

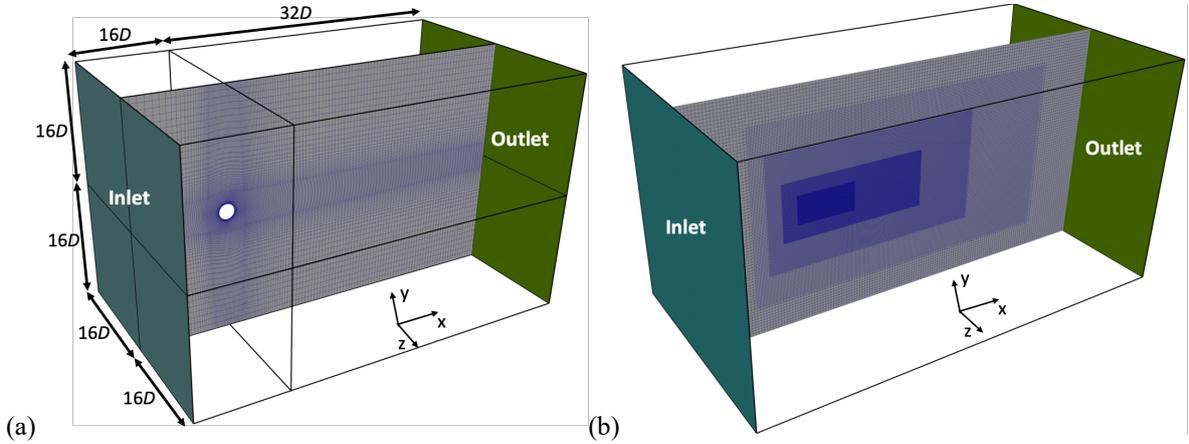

**FIG. 13.** Computational domain and grid distribution around a sphere (a) body-fitted and (b) VPM.

### 2. Steady axisymmetric 3D flow past a sphere

Flow is found to be steady, axisymmetry, and separated from the surface of the sphere for Re in the range of 20 to 210.[21] For the purpose of comparison, the computations are carried out at Re = 100 for both VPM and body-fitted cases. The length of the recirculation region ($L/D$), separation angle from the forward stagnation point ($\theta$) and mean drag coefficient ($Cd_{mean}$) are calculated from the present simulations and compared with the results reported in the literature, as depicted in Table V. The values of $L/D$, $\theta_s$, and $Cd_{mean}$ are in the same range as reported in the literature based on numerical and experimental observations. The maximum errors in the values of $L/D$ and $\theta$ are around 2%, while for $Cd_{mean}$, it is around 5% in comparison to the previous IBM approach.[34]

**TABLE V.** Comparison of results at Re = 100 for flow past a sphere.

| Authors | Technique | $L/D$ | $\theta_s$ | $Cd_{mean}$ |
|---|---|---|---|---|
| Taneda (1956)[50] | Experiment | 0.89 | -- | -- |
| Nakamura (1976)[51] | Experiment | -- | 53 | -- |
| Johnson and Patel (1998)[21] | Simulation | 0.88 | 53 | 1.08 |
| Giacobello *et al*. (2009)[52] | Simulation | 0.88 | 53 | -- |
| Constant *et al*. (2017)[34] | Simulation (IBM OpenFOAM) | 0.92 | 53.03 | 1.14 |
| Present | Simulation (body-fitted) | 0.89 | 53.00 | 1.09 |
| Present | Simulation (VPM) | 0.90 | 53.06 | 1.08 |

### 3. Steady non-axisymmetric 3D flow past a sphere

The flow loses its axial symmetry as Re is increased beyond 210. However, the flow remains steady and exhibits non-axisymmetric behavior for Re in the range of 211 to 270.[21] In order to capture such transitional wake characteristics, the computations are performed at Re = 250, for both VPM and body-fitted cases. The values of the mean drag coefficient ($Cd_{mean}$) and the mean lift coefficient ($Cl_{mean}$) are compared with the numerical results from the literature, as shown in Table VI. These values are in the same range as reported in the literature.[21,34,52]

Figure 14 illustrates the 3D particle path around the sphere in different planar views for both VPM and body-fitted cases. For the purpose of comparison, the results from the literature[21,34] are also included. The first, second, third and fourth columns indicate the particle path from VPM, body-fitted, Constant *et al.*[34] and Johnson and Patel[21], respectively. Figure 14(a) shows the particle path around the sphere in the *x-y* planar view. The fluid coming from the upstream direction forms a spiral in the upper part of the wake region. On the other hand, the lower spiral is formed by the fluid emanating from the upper spiral. The particle path is accurately captured by the present VPM method, indicating its robustness and accuracy. It matches quite well with the present body-fitted case as well as the topology reported by Constant *et al.*[34] and Johnson and Patel[21]. The particle path presented in the *x-z* and *y-z* planes also exhibits similar wake topology as reported in the literature. It can be seen that the flow is asymmetric in the *x-y* view as observed from both VPM and body-fitted cases. The observed flow topology in different views is in good agreement with the results reported in the literature by Constant *et al.*[34] and Johnson and Patel.[21] The three-dimensional streamlines around the sphere projected onto the *x-y* and *x-z* planes are shown in Fig. 15. The 3D streamlines clearly identify the symmetric and asymmetric wake patterns in the *x-z* and *x-y* planes, respectively, as observed in the literature both numerically[21,34] and experimentally[21]. The wake flow behavior from the present VPM method matches quite closely with the body-fitted grid as well as the numerical and experimental results.

Figure 16 (a) and (b) show the distributions of the pressure coefficient in the *x-y* and *x-z* planes obtained in the present VPM and body-fitted simulation at Re = 250, respectively. The pressure contours reported in the previous IBM approach[34] and the numerical study by Johnson and Patel[21] are also shown for comparison. Good agreement in the symmetric and asymmetric pressure distributions in the *x-y* and *x-z* planes can be confirmed among all the cases.

**TABLE VI.** Comparison of results at Re = 250 for flow past a sphere.

| Authors | Technique | $Cl_{mean}$ | $Cd_{mean}$ |
|---|---|---|---|
| Johnson and Patel (1998)[21] | Simulation | -0.061 | 0.70 |
| Giacobello *et al.* (2009)[52] | Simulation | -0.061 | 0.702 |
| Constant *et al.* (2017)[34] | Simulation (IBM OpenFOAM) | -0.062 | 0.72 |
| Present | Simulation (body-fitted) | -0.061 | 0.709 |
| Present | Simulation (VPM) | -0.061 | 0.71 |

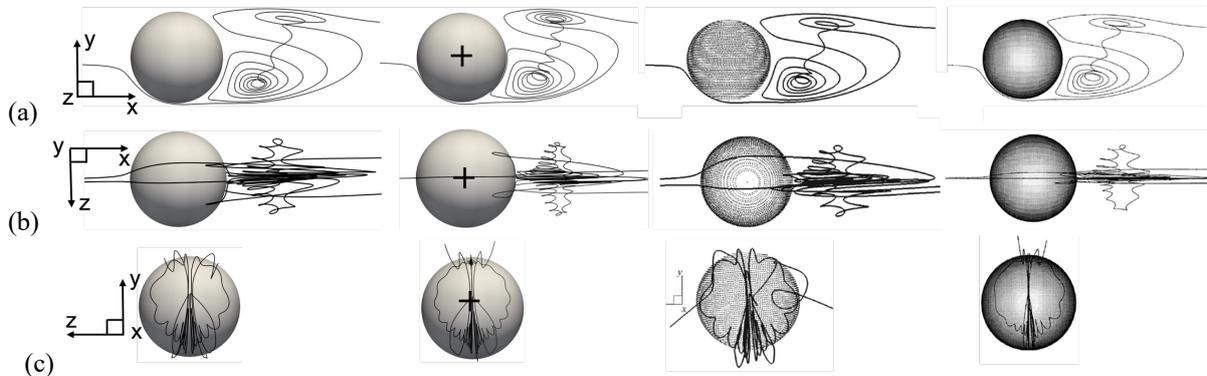

**FIG. 14.** Three-dimensional particle path around the sphere at Re = 250 (a) *x-y* view (b) *x-z* view and (c) *y-z* view. (First column: VPM, second column: body-fitted, third column: Numerical[34] (IBM OpenFOAM), and fourth column: Johnson and Patel[21]).

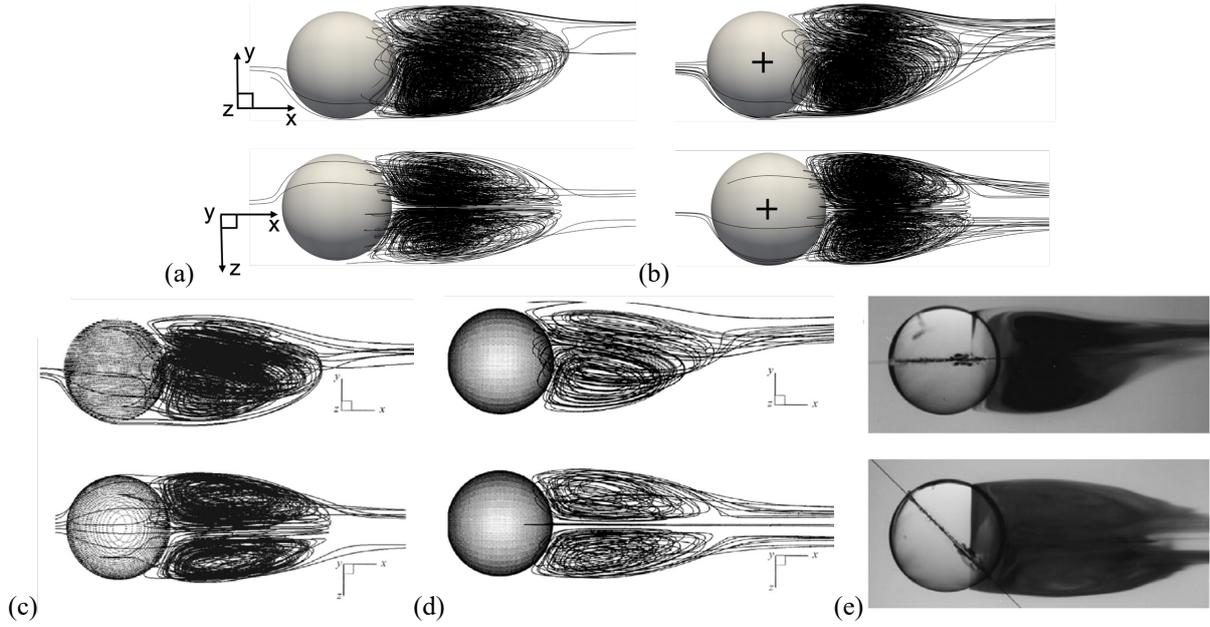

**FIG. 15.** Snapshot of 3D streamlines of *x-y* view (top) and *x-z* view (bottom) flow past a sphere at Re = 250. (a) VPM, (b) body-fitted, (c) Numerical[34] (IBM OpenFOAM) (d) Johnson and Patel[21] (Numerical), and (e) Johnson and Patel[21] (Experiment).

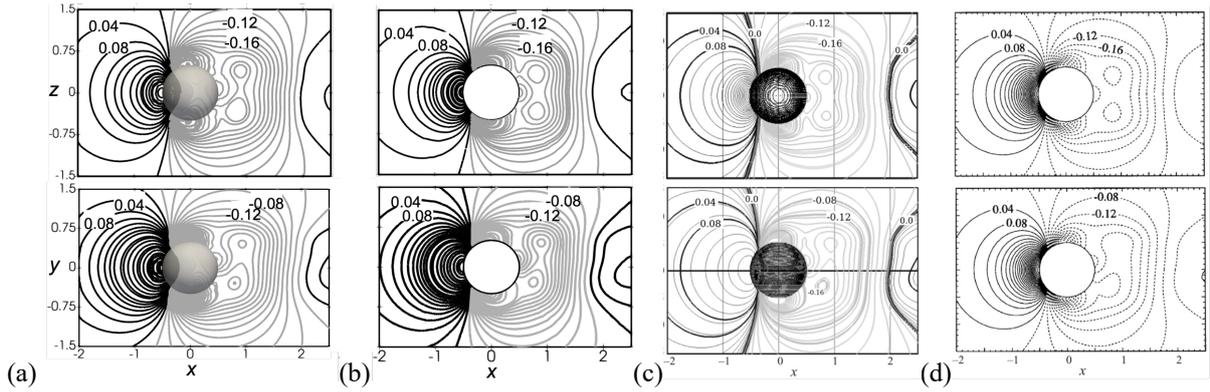

**FIG. 16.** Contours of instantaneous pressure coefficient for a non-axisymmetric flow past a sphere at Re = 250. (a) VPM, (b) body-fitted, (c) Numerical[34] (IBM OpenFOAM), and (d) Johnson and Patel[21] (Numerical) [Top row: *x-z* plane, bottom row: *x-y* plane].

### 4. Unsteady non-axisymmetric 3D flow past a sphere

The transition from steady to unsteady flow occurs as the Re is increased from 270, as reported by Johnson and Patel.[21] The wake structure is observed to be periodic with regular detachment of vortices from the sphere. In order to capture unsteady, periodic, and 3D flow past a sphere, computations are performed at Re = 300 for both VPM and body-fitted cases. The values of $Cd_{mean}$, $Cl_{mean}$, and $St$ are in close agreement with the results reported in the literature, as shown in Table VII. The present VPM results also match closely with the present body-fitted grid.

Figure 17 shows the streamwise mean velocity ($u_{mean}$) and the RMS value of the streamwise velocity fluctuation ($u'_{rms}$) along the streamwise direction. The variation in $u_{mean}$ around the sphere as well as far from the sphere obtained from the present VPM matches quite well with that reported in the previous studies.[21,34] There exist some differences in the present body-fitted and VPM results in the far wake region. This may be primarily due to the coarsening of the grid in the body-fitted case with increasing the distance from the sphere. It should be noted that, however, similar discrepancies can also be observed among the data reported in the previous studies. The streamwise component of the instantaneous velocity ($u$) at each quarter of the periodic oscillation of the wake is plotted along the streamwise axis ($x$) in Fig. 18. The variations in $u$ at different phases in the wake oscillation fall in the same range as reported in the literature. The first peak of the traveling wave is reported to be at $x = 5$ and 4.5 for $t = 0$ by Johnson and Patel[21] and Constant et al.[34], respectively. It is located at $x = 4.75$ in the present VPM and body-fitted cases. The peak of the travelling wake at $t = 0$ moved to $x = 8.5$ and 9 at $t = 3T/4$ ($T$ is the time period of wake oscillation) as reported by Johnson and Patel[21] and Constant et al.[34], respectively. In both the present VPM and body-fitted cases, it also moves to $x = 9$.

**TABLE VII.** Comparison of results at Re = 300 for flow past a sphere.

| Authors | Technique | $Cd_{mean}$ | $Cl_{mean}$ | $St$ |
|---|---|---|---|---|
| Johnson and Patel (1998)[21] | Simulation | 0.656 | 0.069 | 0.136 |
| Giacobello et al. (2009)[52] | Simulation | 0.658 | 0.067 | 0.134 |
| Tomboulides et al. (1993)[53] | Simulation | 0.671 | -- | 0.136 |
| Roos and Willmarth (1971)[24] | Experiment | 0.629 | -- | -- |
| Johnson and Patel (1998)[21] | Experiment | -- | -- | 0.148-165 |
| Constant et al. (2017)[34] | Simulation (IBM OpenFOAM) | 0.679 | 0.066 | 0.139 |
| Present | Simulation (body-fitted) | 0.665 | 0.065 | 0.140 |
| Present | Simulation (VPM) | 0.671 | 0.066 | 0.139 |

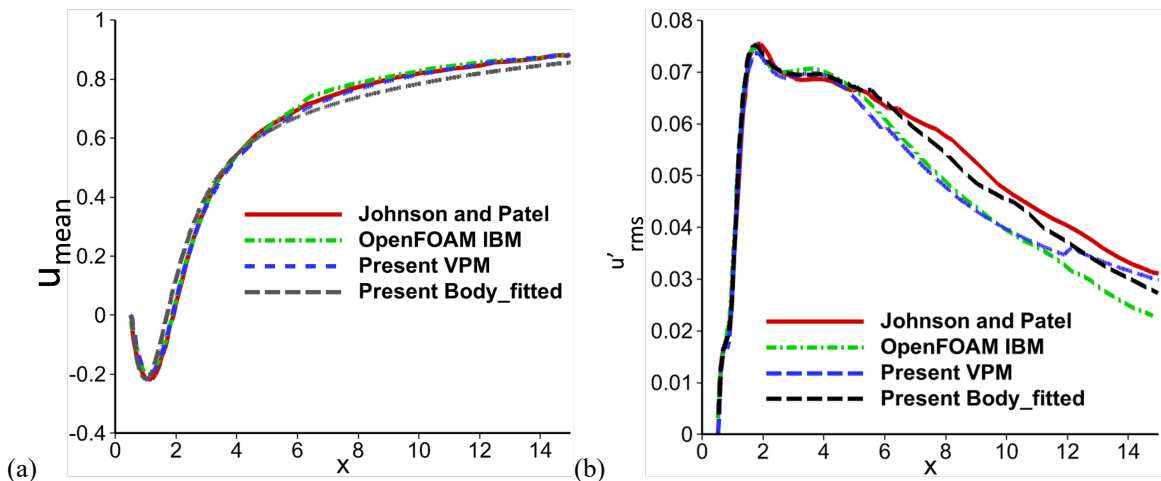

**FIG. 17.** (a) Average streamwise velocity ($u_{mean}$) and (b) rms of the fluctuating component ($u'_{rms}$) along the streamwise direction for flow past a sphere at Re = 300.

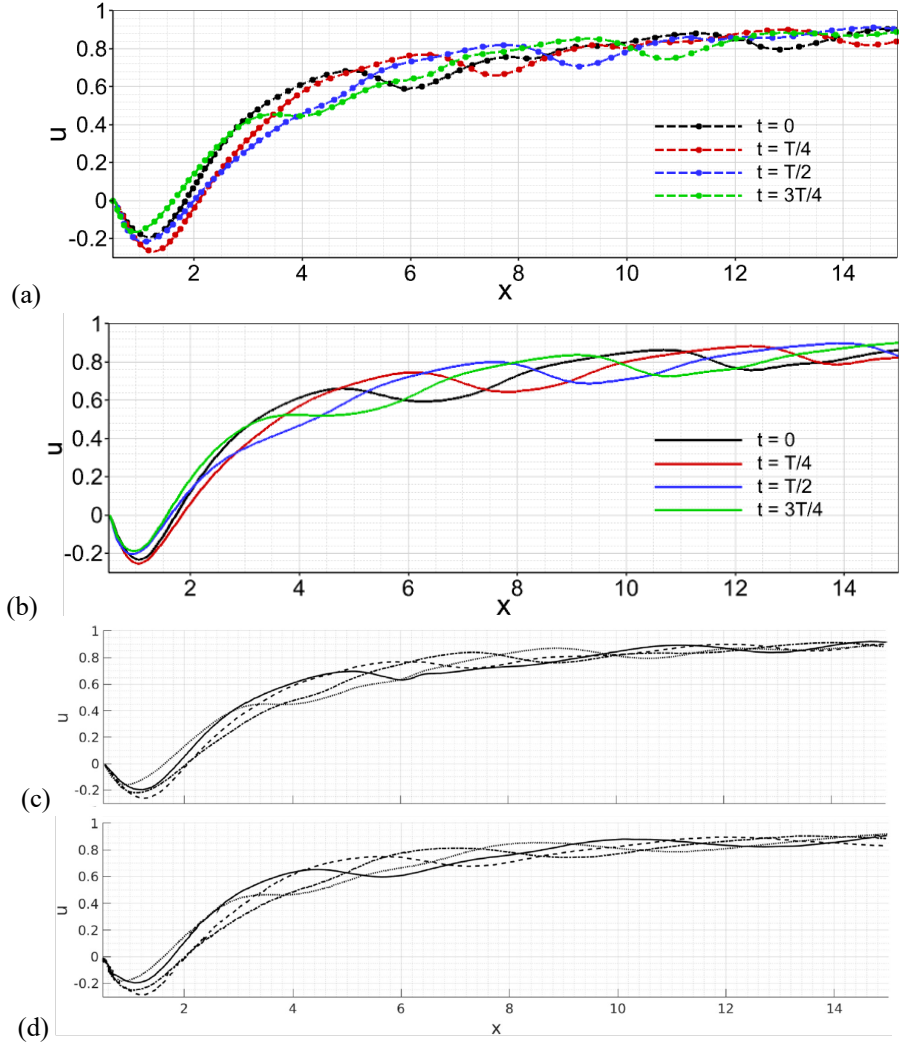

**FIG. 18.** Streamwise velocity (*u*) plotted along the *x* at every quarter period from the rear surface of the sphere. (a) VPM, (b) body-fitted, (c) Numerical[34] (IBM OpenFOAM), and (d) Johnson and Patel[21] (Numerical) [literature: $t = 0$ (–), $t = T/4$ (- -), $t = T/2$ (-.-), $t = 3T/4$ (...)].

## IV. CONCLUSIONS

In the present study, we implemented the volume penalization method, which is a kind of immersed boundary technique, for the first time in OpenFOAM for solving fluid flows past stationary or moving bluff bodies. In this method, an arbitrary three-dimensional shape is embedded into a Cartesian grid system by using a level-set function which is a signed distance function from the fluid-solid interface. Then, the phase indicator is defined so that it possesses the values of 0 and 1 in the fluid and solid regions, respectively, while it is smoothly changed around the fluid-solid interface. The thickness of the transition layer is typically set to be several times the local grid spacing, which can be effectively refined by coupling topoSetDict and refineMeshDict utilities in OpenFOAM..

The solver was verified by considering various benchmark problems in the literature, such as flow past a cylinder and a sphere. Both steady and unsteady characteristics of the wake are compared by considering different global and local parameters. For flow past the cylinder, the Reynolds number (Re) is varied from 40 to 300 which covers steady 2D flow to unsteady 3D flow. In order to further demonstrate the applicability of the present solver to a moving boundary, the flow past a vibrating cylinder in the cross-stream direction at Re = 500 was also considered, and the obtained results show good agreement with those reported in the previous studies. Furthermore, the various transitional regimes, such as steady axisymmetry, steady non-axisymmetry, and unsteady non-axisymmetry flows around a three-dimensional sphere at different Reynolds numbers were also successfully reproduced. The excellent agreement of various mean and temporal quantities, the topological features of the wake, and the response in the force coefficients with previous numerical and experimental studies validate the present VPM developed in the OpenFOAM framework. We have also carried out computations using body-fitted grids employing the same numerical techniques and compared the results with those obtained in the present VPM. The present results indicate that the present VPM is quite promising in simulating flows around bluff bodies with stationary and moving fluid-solid interfaces.

## ACKNOWLEDGEMENT
This study is partially supported by JSPS KAKENHI Grant Numbers JP23H01339, and also Research and Development Program for Promoting Innovative Clean Energy Technologies Through International Collaboration, the New Energy and Industrial Technology Development Organization (NEDO).

## AUTHOR DECLARATIONS
**Conflict of Interest**
The authors have no conflicts to disclose.

**Author Contributions**
**Prashant Kumar**: Conceptualization (equal); Formal analysis (equal); Investigation (equal); Methodology (lead); Software (lead); Validation (lead); Writing–original draft (lead); Writing–review & editing (lead).
**Vivek Kumar**: Conceptualization (equal); Formal analysis (equal); Investigation (equal); Methodology (lead); Software (lead); Validation (lead); Writing–original draft (lead); Writing–review & editing (lead).
**Di Chen**: Conceptualization (equal); Formal analysis (equal); Investigation (equal); Methodology (equal); Software (equal); Validation (equal); Writing–original draft (equal); Writing–review & editing (equal).
**Yosuke Hasegawa**: Conceptualization (equal); Formal analysis (equal); Investigation (equal); Methodology (equal); Supervision (lead); Validation (equal); Writing–original draft (supporting); Writing–review & editing (equal).

## DATA AVAILABILITY
The data that support the findings of this study are available from the corresponding author upon request.